# From Abstract Threats to Institutional Realities: A Comparative Semantic Network Analysis of AI Securitisation in the US, EU, and China

Ruiyi Guo
Beijing Foreign Studies University
202120116130@bfsu.edu.cn

Bodong Zhang
Geneva Graduate Institute
bodong.zhang@graduateinstitute.ch

**Abstract**

*Artificial intelligence governance exhibits a striking paradox: while major jurisdictions converge rhetorically around concepts such as safety, risk, and accountability, their regulatory frameworks remain fundamentally divergent and mutually unintelligible. This paper argues that this fragmentation cannot be explained solely by geopolitical rivalry, institutional complexity, or instrument selection. Instead, it stems from how AI is constituted as an object of governance through distinct institutional logics. Integrating securitisation theory with the concept of the dispositif, we demonstrate that jurisdictions govern ontologically different objects under the same vocabulary. Using semantic network analysis of official policy texts from the European Union, the United States, and China (2023–2025), we trace how concepts like safety are embedded within divergent semantic architectures. Our findings reveal that the EU juridifies AI as a certifiable product through legal-bureaucratic logic; the US operationalises AI as an optimisable system through market-liberal logic; and China governs AI as socio-technical infrastructure through holistic state logic. We introduce the concept of structural incommensurability to describe this condition of ontological divergence masked by terminological convergence. This reframing challenges ethics-by-principles approaches to global AI governance, suggesting that coordination failures arise not from disagreement over values but from the absence of a shared reference object.*

*Keywords:* AI governance; securitisation; semantic network analysis; institutional logics; dispositif; structural incommensurability; comparative policy analysis

## 1. Introduction and Research background

Artificial intelligence has rapidly shifted from a relatively peripheral technical issue to a central object of public authority and political concern. Over the past decade, AI has been framed not only as a driver of economic growth and technological innovation, but increasingly as a source of systemic risk, with potentially far-reaching implications for social order, economic structures, and political governance. This transformation is commonly described as a process of securitisation, whereby algorithmic systems are elevated from ordinary policy domains into objects requiring priority, and sometimes urgent, intervention, thus entering the core agenda of national and transnational governance (Cihon, Maas and Kemp, 2020).

This shift is evident across the three most influential jurisdictions in contemporary AI development and regulation. The European Union has adopted the world's first comprehensive and binding regulatory framework in the form of the *AI Act*, explicitly defining AI systems as technologies capable of infringing fundamental rights and embedding security governance through ex ante risk classification and conformity assessment (Veale and Borgesius, 2021; Bradford, Aboy and Liddell, 2023). In the United States, AI governance has largely been constructed through risk management frameworks that emphasise the identification, assessment, and mitigation of harms in specific application contexts, while preserving substantial scope for industry self-regulation and innovation (Kroll et al., 2017; National Institute of Standards and Technology, 2023). In contrast, AI governance in China is closely integrated into broader strategies of state development and social governance, where security objectives are intertwined with social stability, information control, and technological self-reliance, producing a more holistic governance logic (Creemers, 2018; Ding, 2018;

Roberts et al., 2021).

Despite these divergent regulatory trajectories, political discourse across jurisdictions exhibits striking surface convergence. Policy documents increasingly rely on a shared vocabulary of governance, including terms such as *safety*, *risk*, *transparency*, *accountability*, and *responsibility*. Events such as the 2023 Bletchley Park AI Safety Summit have reinforced the impression that a global consensus on AI governance principles is emerging. Accordingly, a significant strand of the literature conceptualises global AI governance primarily as a coordination problem, suggesting that regulatory divergence results mainly from institutional frictions, strategic competition, or enforcement costs, while consensus at the level of principles has largely been achieved (Floridi et al., 2018; Cihon et al., 2021; Schmitt, 2022).

Recent comparative research and regulatory practice, however, challenge this assumption. Even among jurisdictions that explicitly endorse similar governance principles, AI regulation diverges sharply in scope, enforcement mechanisms, and policy priorities (Cath, 2018; Marsden and Meyer, 2023). Luna et al. (2024) demonstrate that while countries converge rhetorically around concepts such as safety and risk, this convergence remains confined to abstract goals and terminology. Once governance enters concrete regulatory procedures and enforcement mechanisms, coherence rapidly dissolves, resulting in terminological convergence coupled with operational fragmentation. This raises a central puzzle: if AI has been successfully securitised across major jurisdictions, why has this shared sense of political urgency failed to generate comparable governance architectures, and why do governance outcomes remain mutually unintelligible despite extensive overlap in regulatory vocabulary?

Existing theories offer partial explanations for this fragmentation. Realist approaches emphasise geopolitical rivalry and strategic interests, particularly between the United States and China, treating AI as a strategic asset whose governance is shaped by competition for economic and military advantage (Horowitz, 2018; Allen and Chan, 2023). Institutional accounts focus on regime complexity and coordination costs arising from overlapping institutions and the absence of a central coordinating authority (Raustiala and Victor, 2004; Cihon, Maas and Kemp, 2020). What's more, scholarship on policy networks and regulatory instruments highlights differences in legal tools, standards, and market mechanisms (Abbott et al., 2017; Black, 2008; Gunningham, 2021). While these perspectives illuminate important dynamics, they share a common limitation: they implicitly assume that jurisdictions are governing the same underlying object, differing only in interests, institutions, or instruments.

This paper argues that fragmentation in AI governance cannot be fully understood without examining how AI itself is constituted as an object of governance. By integrating securitisation theory with institutional logics through the lens of the dispositif, the paper shows that AI securitisation is operationalised through pre-existing institutional logics that shape how threats are defined, stabilised, and rendered governable. Jurisdictions may converge rhetorically on "AI safety" while governing fundamentally different things.

To substantiate this claim, the paper combines securitisation theory and the dispositif perspective with semantic network analysis of official policy texts from the European Union, the United States, and China. Using topological indicators such as degree, betweenness, and eigenvector centrality, alongside semantic neighbourhood analysis based on co-occurrence patterns, the study traces how concepts such as safety and risk are embedded within distinct semantic architectures. Rather than treating shared terminology as evidence of convergence, the analysis reveals how governance priorities are structured through divergent institutional logics, producing fragmented and ontologically incommensurable AI governance regimes.

## 2. Theoretical Framework and Hypothesis

To explain the paradox of rhetorical convergence and operational fragmentation in global AI governance, this study integrates securitisation theory and institutional logics, interpreted through Michel Foucault's concept of the dispositif. Securitisation theory explains how artificial intelligence is elevated to the level of high politics, as political actors construct AI as an existential or systemic threat through discursive and symbolic practices (Buzan, Wæver and de Wilde, 1998; Cihon, Maas and Kemp, 2020). Institutional logics, by contrast, explain why these securitised threats are translated into divergent governance practices across administrative and bureaucratic contexts. The concept of the dispositif links these perspectives by capturing the material and procedural mechanisms through which abstract security rationalities are enacted.

### 2.1 Securitisation and Institutional Logics

Securitisation theory holds that security is not an objective condition, but a socially constructed status that emerges when actors successfully frame a phenomenon as an existential threat requiring extraordinary measures (Buzan, Wæver and de Wilde, 1998). In the context of artificial intelligence, narratives of existential and systemic risk function as securitising moves that mobilise political attention, institutional authority, and regulatory resources. While securitisation theory effectively explains why AI has entered the security agenda, it offers limited insight into how securitised threats are operationalised within different governance systems.

Institutional logics provide a framework for understanding this variation. Defined as historically sedimented patterns of practices, assumptions, and rules that structure organisational behaviour (Friedland and Alford, 1991; Thornton, Ocasio and Lounsbury, 2012), institutional logics supply the cognitive and normative grammar through which abstract threats are translated into routine governance practices. In the contexts examined here, distinct institutional logics generate different interpretations of AI-related threats. In the United States, a market-liberal logic frames threats as probabilistic risks that can be managed across sectors through assessment, mitigation, and market-based regulation. In the European Union, a legal-bureaucratic logic interprets threats as potential infringements on rights and legal order, requiring ex ante classification and compliance procedures. In China, a holistic state logic treats threats as challenges to social stability and national development, generating comprehensive, state-centred regulatory responses. Although the impulse to securitise AI is global, the concrete implementation of security is therefore path-dependent and embedded within local institutional logics.

### 2.2 Dispositif and Governable Unites

While institutional logics offer cognitive templates, abstract threat judgments require concrete mechanisms to become administrative action. This operational dimension is captured through the concept of the *dispositif*. Foucault defines this as a heterogeneous ensemble comprising discourses, laws, and technical practices that responds to strategic historical urgencies (Foucault, 1980: 194-195). Rose and Miller (1990, 1992, 2008) refine this framework by distinguishing between *political rationalities*, which render objects thinkable, and *technologies* *of government*, which make intervention possible. The latter, encompassing arrangements such as classification standards and compliance procedures, operationalise political rationalities (Rose and Miller, 1990: 8).

Crucially, technologies of government transform complex realities into discrete *governable units*. Abstract goals like safety must be translated into specific frameworks. For instance, the EU's high-risk system defined by rights impact, China's model service focused on information content, and the US automated system identified by substantive intervention represent distinct units produced by divergent dispositif configurations. This theoretical framework underpins the methodological approach of this study. Since language inscribes reality into governance calculations (Rose and Miller, 1990: 7), the organisation of concepts in policy texts, specifically their centrality and systematic associations, represents the structural traces of the dispositif. Aligning with the materialist turn in AI governance (Gordon, Rieder and Sileno, 2022: 2), these semantic patterns are treated as identifying markers of governance logic.

Consequently, Semantic Network Analysis is employed to trace these discursive footprints. By analysing topological metrics, the characteristics of *dispositif* configurations are indirectly mapped across jurisdictions. This approach allows for testing whether institutional logics fundamentally shape the organisation of governance discourse and provides a robust pathway for systematic cross-national comparison.

### 2.3 Hypothesis

Building on the premise that institutional logics provide the grammar through which abstract threats are translated into routine governance practices (Friedland and Alford, 1991), this study hypothesises that the semantic topologies of AI safety will reflect the distinctive political rationalities of each jurisdiction. Three hypotheses are advanced and operationalised through network topology and semantic neighbourhood analysis.

**H1: In the European Union, the dominance of a legal-bureaucratic logic is expected to produce a network structure characteristic of horizontal product governance.**

Governance prioritises rule codification and the protection of fundamental rights, framing AI threats as potential infringements on legal order. Accordingly, the network is expected to exhibit high degree centrality around legal classification terms, particularly *high_risk*. Procedural nodes such as *conformity_assessment, compliance* are expected to display high eigenvector centrality, functioning as the executive core that links regulatory definition to market access. In semantic neighbourhoods, safety is expected to co-occur closely with established product liability vocabulary, such as *fundamental_rights, market* and *criteriation* .

**H2: In the United States, a market-liberal logic is expected to generate a topology of sectoral risk management.**

Safety is constructed as a probabilistic variable optimised against innovation costs rather than a binary condition of legality. Consequently, risk is expected to replace safety as the primary anchoring concept. Executive influence, which is measured through eigenvector centrality, is expected to be distributed across downstream sectors such as *finance* and *employment*. In semantic neighbourhoods, safety is expected to be closely associated with technical performance metrics such as *robustness* and *reliability*, alongside terms describing market externalities like *harm* and *bias*.

**H3: In China, a holistic state logic is expected to produce a topology of full-stack stability governance.**

This logic prioritises political order, treating development and security as inseparable. The term *anquan (Safety/Security)* is therefore expected to function as a super-node encompassing both technical safety and national security. The network's executive core is expected to be defined by content-oriented concepts such as *generative_content (shengcheng neirong)* rather than system architecture. In semantic neighbourhoods, *anquan* is expected to co-occur closely with *stability (wending)* and *society (shehui)*, indicating a governance focus on maintaining social order through the regulation of information outputs.

## 3. Empirical design and methodology

### 3.1 Corpus Construction and Preprocessing

A stratified corpus of authoritative policy documents was constructed for the European Union, the United States, and China, covering the critical regulatory period from 2023 to 2025. The corpus comprises legislative texts (e.g., the *EU AI Act*), executive instruments (e.g., *US Executive Order 14110*), and strategic frameworks (e.g., China's *Global AI Governance Initiative*).

Given the linguistic heterogeneity of the source texts, a uniform pre-processing pipeline was implemented in Python to ensure cross-jurisdictional comparability. Raw texts were extracted via pdfplumber and subjected to regex-based cleaning to remove digitisation artefacts. Crucially, to preserve the integrity of governance concepts, a dictionary-based n-gram recognition procedure was applied. Domain-specific multi-word expressions (MWEs) such as *high_risk, generative_synthesis* (*shengcheng hecheng*), and *conformity_assessment* were fused into single tokens. For the Chinese sub-corpus, segmentation was performed using a hybrid approach combining jieba tokenisation with a custom dictionary of 200+ policy terms, ensuring that compound concepts (e.g., *quanyi*, "rights/interests") were treated as distinct semantic units equivalent to their English counterparts.

### 3.2 Semantic Network Construction

Latent structural relationships were modelled by constructing undirected and weighted semantic networks for each jurisdiction. Unlike topic modelling approaches that cluster documents based on word distributions, this study maps the topology of concept association. Associations were identified using a sliding window approach with a window size of 15 tokens, which was calibrated to approximate the semantic span of a complex policy clause. A hybrid edge-weighting algorithm was developed to filter syntactic noise while preserving the structural backbone of the discourse. First, Normalised Pointwise Mutual Information (NPMI) was computed for all co-occurring pairs. Edges with low NPMI values were pruned to remove trivial associations driven purely by high frequency. Second, to prevent the fragmentation of essential anchor terms which are often penalised by NPMI due to their ubiquity, a back-off mechanism retained edges for designated protected nodes such as *ai* and *safety* based on raw co-occurrence counts if they exceeded a minimum threshold. This hybrid construction ensures that the resulting networks reflect both statistically significant semantic bindings and the robust structural skeleton of the discourse.

### 3.3 Metrics Operationalisation

Three topological metrics were calculated to operationalise the components of the *dispositif* as summarised in the subsequent analysis. First, degree centrality was calculated to measure the absolute connectivity of a node. High-degree nodes are interpreted as the primary anchors of problematisation, indicating what the system views as the central object of governance. Second, eigenvector centrality was employed to measure the influence of a node based on its connections to other high-scoring nodes. In this study, eigenvector centrality serves as a proxy for the executive core of the network. It identifies the mechanisms, such as *conformity_assessment* in the EU or *finance* in the US, that serve as the functional passage points for exercising power. Finally, the semantic neighbourhoods of key contested concepts like *safety* or *anquan* were analysed using NPMI ranking. Unlike raw frequency lists, NPMI isolates terms that share a specific and non-random semantic affinity. This metric allows for the qualitative distinction between safety framed as *health* in a product logic, *robustness* in an engineering logic, or *stability* in a political logic.

The comparative semantic network analysis of the EU, US, and Chinese corpora reveals a significant structural divergence beneath the surface of a shared global lexicon. While policy documents across these jurisdictions frequently employ identical terms, the topological organisation of these concepts differs substantially. As summarised in Table 1, distinct patterns in degree centrality, eigenvector centrality, and semantic neighbourhoods indicate that institutional logics play a constitutive role in translating abstract security concerns into concrete governance mechanisms. These findings suggest that the three jurisdictions are not merely applying different regulations to the same phenomenon but are governing ontologically distinct objects defined by their specific institutional contexts.

## 4. Results and Discussion

| Dimension | EU | US | China |
|---|---|---|---|
| Problematisation Anchor (Top5 Degree Centrality) | high_risk (0.367) | risk (0.388) | 安全 (Security&Safety) (0.286) |
| | ai_system (0.327) | ai (0.367) | 人工智能(AI) (0.245) |
| | right (0.306) | right (0.367) | 模型 (model) (0.225) |
| | fundamental_rights (0.306) | privacy (0.347) | 探索 (exploration) (0.163) |
| | trustworthy_ai(0.286) | automated_system (0.327) | 服务提供商(service provider) (0.163) |
| Execution Core (Top5 Eigenvector Centrality) | conformity_assessment (0.455) | education(0.437) | 生成内容(generative synthesis) (0.484) |
| | notified_body (0.451) | employment (0.401) | 标注 (marking/labelling) (0.460) |
| | quality_management (0.451) | finance (0.401) | 内容 content (0.421) |
| | technical_documentation (0.399) | justice (0.385) | 标识 (identifiers) (0.402) |
| | documentation (0.243) | health (0.357) | 服务提供商 (service provider) (0.345) |
| Safety Semantics (High 5 NPMI with *Safety*/*Anquan*) | health (0.517) | robustness (0.383) | 稳定 (stability) (0.242) |
| | security (0.319) | reliability (0.342) | 测评(evaluation/testing) (0.226) |
| | trustworthy_ai (0.270) | risk_mitigation (0.225) | 权益 (rights/interests) (0.225) |
| | information_security (0.202) | auditor (0.225) | 威胁性 (threateningness) (0.214) |
| | internal_market (0.194) | freedom (0.216) | 备案 (filing) (0.196) |

Table 1: Comparative Topological Metrics of AI Governance Discourse across the EU, US, and China (2023–2025)

### 4.1 The European Union: The Bureaucratic

**Passage Point (Testing H1)**

The semantic network derived from the EU policy corpus (Figure 1) reveals a densely structured topology that closely mirrors the characteristics of a legal-bureaucratic institutional logic. Unlike a decentralized risk network, the EU's structure is organised around rigid classification nodes and procedural checkpoints, confirming the expectation of horizontal product governance.

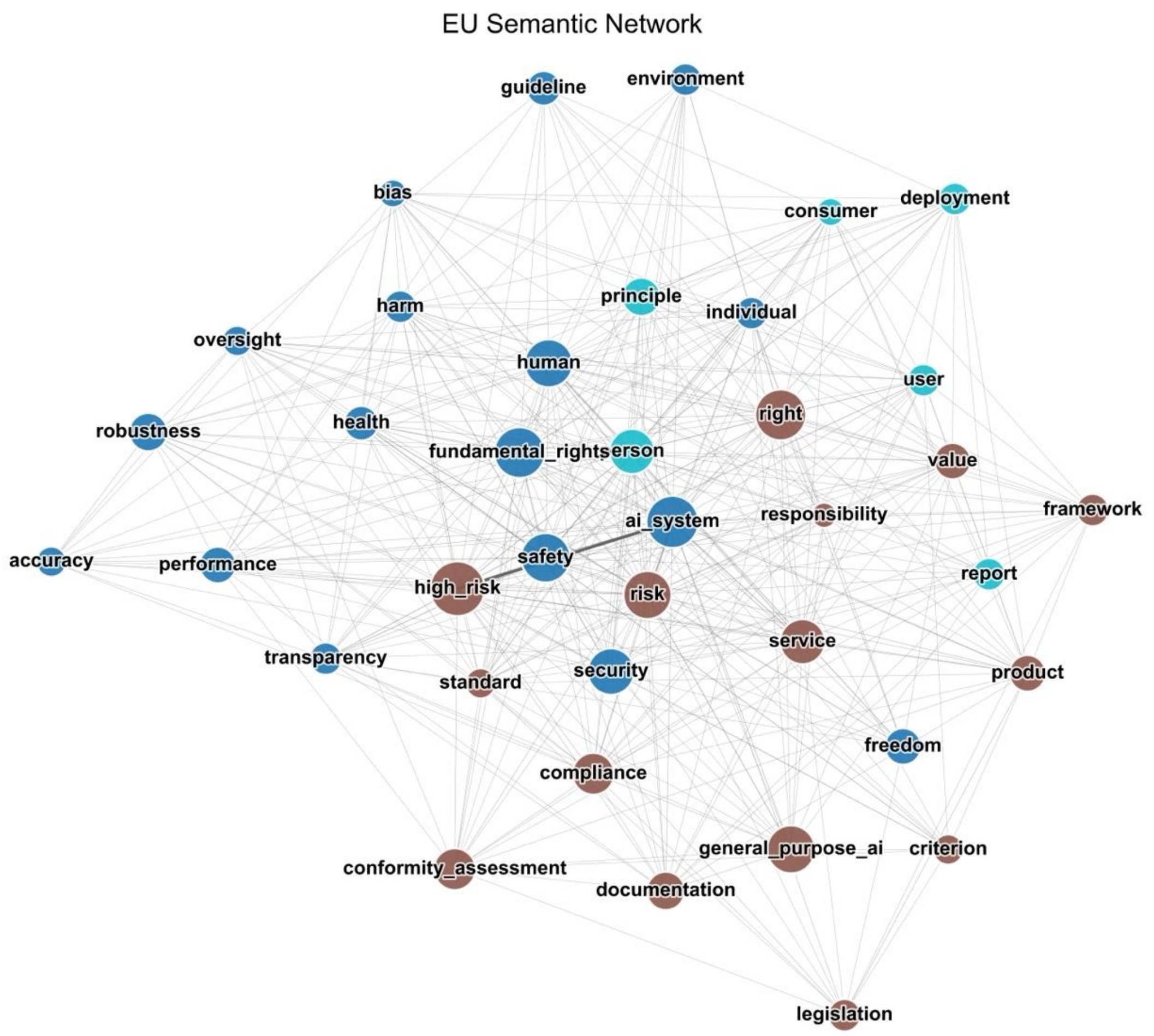

Figure 1: The European Union Semantic Network: A Topology of Horizontal Product Governance

As detailed in Table 1, the network's problematisation logic is anchored by *high_risk* (Degree Centrality = 0.367, see appendix A for full metrics). Although not the most frequent token in absolute terms, its high centrality indicates its function as a strategic pivot: in the EU's dispositif, high_risk is not merely a descriptive label but a legal trigger that activates the entire regulatory apparatus. This node is structurally coupled with *right* (0.306) and *fundamental_rights* (0.306), situating AI governance firmly within the constitutional acquis of the Union rather than treating it as a novel, standalone technical domain.

The execution core of the network, measured by Eigenvector Centrality in Table 1, is dominated by procedural mechanisms rather than substantive outcomes. The highest-ranking nodes, *conformity_assessment* (0.455), *notified_body* (0.451), and *quality_management* (0.451), together form a tight bureaucratic passage point. This structural configuration suggests that governance authority is exercised through the verification of documentation and process compliance. The network effectively "locks" AI systems into a certification chain, where market access is conditional upon passing through these administrative nodes.

| Region | Anchor | Neighbor | NPMI | Cooc_Windows | Anchor_WindowCoun | Neighbor_WindowCoun |
|---|---|---|---|---|---|---|
| EU | Safety | health | 0.517 | 1626 | 975 | 715 |
| | | security | 0.3192 | 1626 | 866 | 460 |
| | | trustworthy_artificial_intelligence | 0.2698 | 1626 | 67 | 55 |
| | | ai_systems | 0.2451 | 1626 | 15 | 15 |
| | | other_non-personal_data | 0.217 | 1626 | 15 | 13 |
| | | high_risk | 0.2019 | 1626 | 15 | 12 |
| | | information_security | 0.2019 | 1626 | 15 | 12 |
| | | fundamental_rights | 0.1989 | 1626 | 1559 | 568 |
| | | internal_market | 0.1936 | 1626 | 150 | 79 |
| | | technical_documentation | 0.1859 | 1626 | 15 | 11 |
| | | risk | 0.1803 | 1626 | 2661 | 858 |
| | | ai_system_output | 0.1688 | 1626 | 15 | 10 |
| | | high_risk | 0.1688 | 1626 | 15 | 10 |
| | | robustness | 0.1609 | 1626 | 341 | 141 |
| | | threat | 0.1428 | 1626 | 170 | 72 |
| | | notified_body | 0.1306 | 1626 | 15 | 8 |
| | | systemic_risk_producing | 0.1306 | 1626 | 15 | 8 |
| | | risk_assessment | 0.1215 | 1626 | 250 | 93 |
| | | product | 0.1212 | 1626 | 990 | 317 |
| | | criterion | 0.1204 | 1626 | 372 | 132 |

Table 2: Semantic Neighborhood of the 'Safety' Anchor in the EU Corpus (Ranked by NPMI)

The semantic neighbourhood analysis of the safety anchor (Table 2) provides the most direct evidence for H1. In the EU context, safety is most strongly associated with *health* (NPMI = 0.517), followed by *security* (0.319). This association with health is particularly revealing: it indicates that AI safety is conceptualised through the lens of established product safety directives, akin to medical devices or machinery, where safety is defined as the absence of physical injury or adverse health impacts. Crucially, procedural terms like conformity_assessment do not appear in this immediate neighbourhood, implying a functional division: safety serves as the normative goal anchored in health and rights, while the bureaucratic machinery functions as the distinct mechanism to enforce it.

Taken together, these findings strongly support H1. The EU does not govern AI as a dynamic socio-technical risk to be optimised, but as a stable "high-risk product" that should be certified. The governance logic is one of *ex ante* admissibility, where the primary objective is to ensure that the technical artefact fits within the existing legal ontology of the Single Market.

### 4.2 The United States: The Sectoral Risk Matrix (Testing H2)

The United States' semantic network (Figure 2) displays a decentralised topology that aligns with a market-liberal logic, privileging sectoral risk management over centralised legal control. In direct contrast to the EU, Table 1 shows that *risk* (0.388) acts as the primary problematisation anchor, whereas *high_risk* appears only as a peripheral node. This structural inversion suggests that governance focuses on managing probabilistic outcomes rather than establishing binary legal categories.

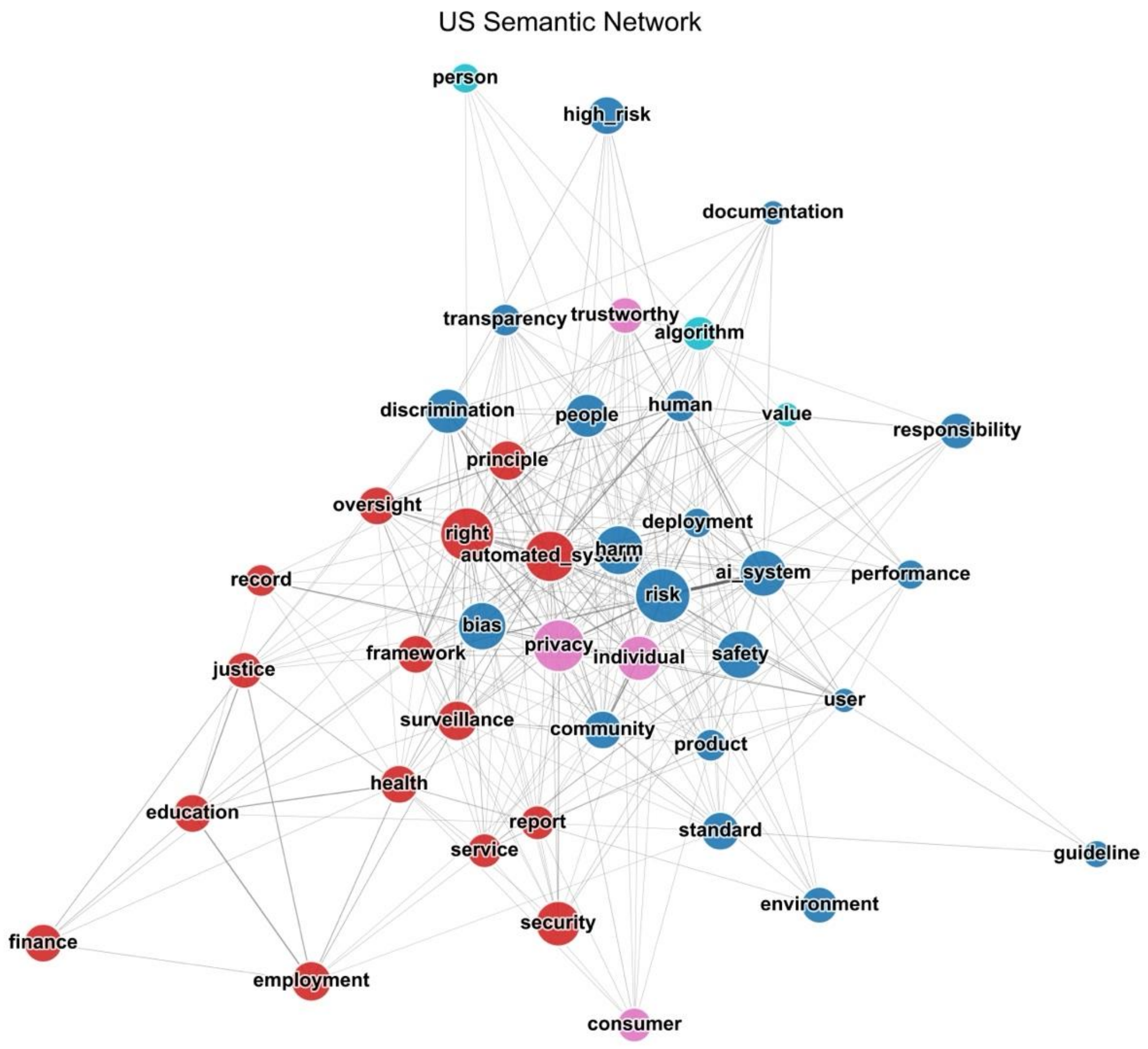

Figure 2: The United States Semantic Network: A Topology of Sectoral Risk Management

The execution core of the network reflects this dispersed authority. As shown in Table 1, eigenvector centrality is distributed across downstream application domains such as *education* (0.437), *employment* (0.401), and *finance* (0.401). Unlike the EU's bureaucratic bottleneck, the US topology embeds governance within specific sectors, indicating a model where regulatory responsibility is delegated to domain-specific agencies and market mechanisms. This confirms the expectation that US governance operates as a sectoral risk matrix, where the definition of acceptable risk varies according to the specific context of deployment.

| Region | Anchor | Neighbor | NPMI | Cooc_Windows | Anchor_WindowCoun | Neighbor_WindowCoun |
|---|---|---|---|---|---|---|
| US | Safety | robustness | 0.3831 | 280 | 15 | 14 |
| | | reliability | 0.3415 | 280 | 15 | 12 |
| | | risk_mitigation | 0.2253 | 280 | 30 | 13 |
| | | auditor | 0.2253 | 280 | 30 | 13 |
| | | freedom | 0.2163 | 280 | 53 | 20 |
| | | ai_ethic | 0.2061 | 280 | 30 | 12 |
| | | product | 0.1923 | 280 | 164 | 47 |
| | | harm | 0.1815 | 280 | 657 | 147 |
| | | platform | 0.1783 | 280 | 45 | 15 |
| | | security | 0.1541 | 280 | 285 | 67 |
| | | service | 0.1477 | 280 | 213 | 51 |
| | | deployment | 0.1249 | 280 | 406 | 84 |
| | | decision_making | 0.1018 | 280 | 120 | 26 |
| | | accuracy | 0.0974 | 280 | 62 | 14 |
| | | high_risk | 0.0931 | 280 | 109 | 23 |
| | | person | 0.0927 | 280 | 162 | 33 |
| | | machine_learning | 0.092 | 280 | 120 | 25 |
| | | compliance | 0.0884 | 280 | 75 | 16 |
| | | principle | 0.0744 | 280 | 581 | 101 |
| | | employment | 0.0721 | 280 | 240 | 44 |

Table 3: Semantic Neighborhood of the 'Safety' Anchor in the US Corpus (Ranked by NPMI)

The semantic neighbourhood of safety (Table 3) offers critical insight into how this logic operationalises protection. The strongest associations are found not with rights, but with technical performance metrics: *robustness* (NPMI = 0.383) and *reliability* (0.342). This finding is significant as it indicates that safety is constructed as an engineering property of the product, which is a measure of how well a system functions under stress, rather than a protective legal status. Furthermore, the presence of *risk_mitigation* (0.225) and *harm* (0.182) in the immediate neighbourhood suggests a reactive governance model. Safety interventions are triggered by specific, demonstrable harms resulting from system failures, rather than by the inherent classification of the technology.

These patterns confirm H2. US governance functions by converting abstract threats into technical problems of optimization. Safety is treated as a contingent variable determined by technical reliability (*robustness*) and context-specific cost–benefit assessments, rather than as a constitutional guarantee. The governable object here is not a "high-risk product" but a "reliable system: whose externalities must be managed through market corrections and sectoral oversight.

### 4.3 China: The Logic of Full-Stack Stability (Testing H3)

The semantic network derived from the Chinese policy texts (Figure 3) exhibits a topology of full-stack stability governance, characterised by the seamless integration of technical administration and political control. Unlike the segmented or sector-specific structures observed in the EU and US, the Chinese network is cohesively organised around the super-node *anquan* (Safety&Security).

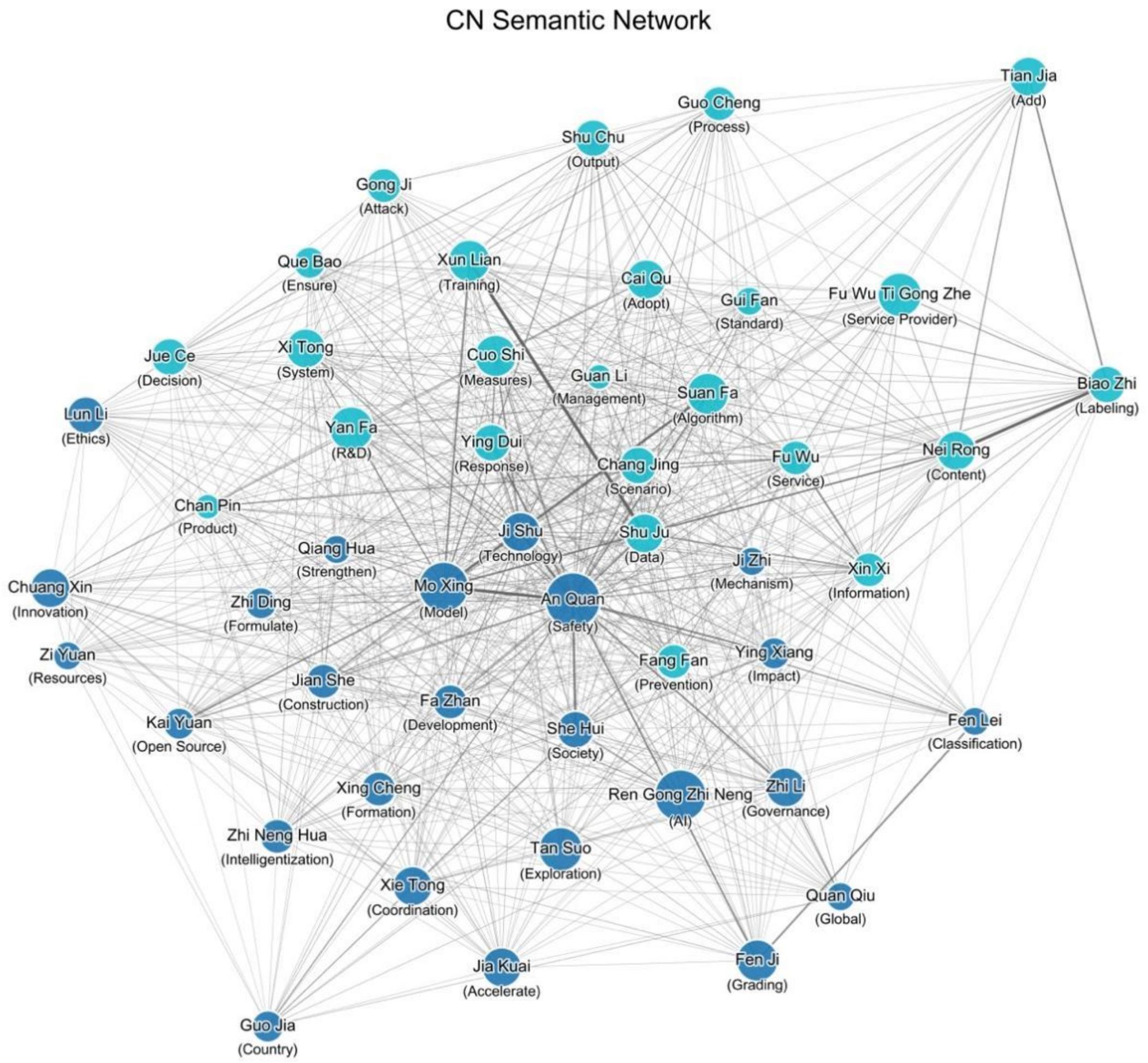

Figure 3: The Chinese Semantic Network: A Topology of Full-Stack Stability Governance

As indicated in Table 1, while the degree centrality of *anquan* (0.286) is quantitatively lower than the US risk anchor, its topological role is distinct: it functions as a comprehensive bridge. It is the sole node that connects technical layers (*model, data, algorithm*) directly with macro-political objectives (*society, governance*). This structure supports the "holistic state logic" described in H3, where distinctions between technical safety and national security are dissolved.

The execution core of the network, measured by eigenvector centrality in Table 1, reveals a unique focus on information control. The most influential nodes are not legal procedures or market sectors, but content-specific technical terms, such as *generative_synthesis* (0.484) and *content* (0.421). This suggests that the primary governable unit is the information stream itself. Governance authority is exercised by penetrating the technical value chain to ensure that algorithmic outputs align with state priorities.

| Region | Anchor | Neighbor | NPMI | Cooc_Windows | Anchor_WindowCoun | Neighbor_WindowCoun |
|---|---|---|---|---|---|---|
| CN | 安全 (safety&securty) | 稳定 (stability) | 0.2424 | 128 | 2294 | 135 |
| | | 测评 (evaluation/testing) | 0.226 | 99 | 2294 | 105 |
| | | 权益 (rights/interest) | 0.2251 | 75 | 2294 | 75 |
| | | 威胁性 (threatingness) | 0.2143 | 60 | 2294 | 60 |
| | | 原则 (principles) | 0.2024 | 102 | 2294 | 120 |
| | | 释放 (release) | 0.2018 | 45 | 2294 | 45 |
| | | 潜在 (potential) | 0.2011 | 112 | 2294 | 135 |
| | | 链条 (chain) | 0.201 | 57 | 2294 | 60 |
| | | 分级 (grading) | 0.1992 | 191 | 2294 | 258 |
| | | 威胁 (threat) | 0.1975 | 89 | 2294 | 104 |
| | | 备案 (filing) | 0.1963 | 44 | 2294 | 45 |
| | | 渐进 (progressive) | 0.1963 | 44 | 2294 | 45 |
| | | 安全意识 (security awareness) | 0.1919 | 55 | 2294 | 60 |
| | | 社会 (society) | 0.1892 | 357 | 2294 | 562 |
| | | 标准规范 (standards and specification | 0.1873 | 54 | 2294 | 60 |
| | | 基础 (basis) | 0.1864 | 30 | 2294 | 30 |
| | | 级别 (level) | 0.1864 | 30 | 2294 | 30 |
| | | 严重 (severity) | 0.1864 | 30 | 2294 | 30 |
| | | 定级 (rating) | 0.1864 | 30 | 2294 | 30 |
| | | 消减 (mitigation) | 0.1864 | 30 | 2294 | 30 |

Table 4: Semantic Neighborhood of the "Anquan"(Safety/Security) Super-Node in the Chinese Corpus

The semantic neighbourhood analysis in Table 4 provides the final empirical confirmation of this holistic logic. The strongest association is found with *stability* (*wending*, NPMI = 0.242), followed closely by *evaluation* (*ceping*, 0.226) and *rights/interests* (*quanyi*, 0.225). The dominance of stability confirms that in the Chinese context, *anquan* is operationally synonymous with the maintenance of social order, differing fundamentally from the US focus on technical robustness or the EU focus on product safety.

More importantly, the data reveals a dual-focus in governance mechanisms. On one hand, the high ranking of *rights/interests* (0.225) and *society* (*shehui*, 0.189) suggests that legitimacy is constructed by framing stability as a prerequisite for protecting collective social interests. On the other hand, the cluster of procedural terms, such as *evaluation* (0.226), *grading* (*fenji*, 0.199), and *filing* (*bei'an*, 0.196), point to a hands-on, *governance-by-architecture* approach. Unlike the post-market enforcement seen elsewhere, these terms indicate that the state actively intervenes in the technical lifecycle through continuous testing, risk grading, and administrative registration.

In conclusion, China's governance regime constitutes a full-stack intervention where AI is treated as a critical socio-technical infrastructure, which support the H3. By binding stability to society and enforcing it through filing and grading, the state ensures that the technology's development and output, from training data to generated content, remain continuously stabilised within the political framework.

### 4.4 Discussion: From Divergent Topologies to Structural Incommensurability

The semantic network data presented in sections 4.1 through 4.3 point to a central question regarding the nature of global alignment. While policy texts across the three jurisdictions share a global vocabulary of *safety* and *risk*, they produce starkly different network topologies. A superficial reading might suggest that this divergence merely reflects differing governance preferences or regulatory styles. However, the topological evidence reveals a more fundamental schism where each jurisdiction constructs AI as a distinct type of governable object.

Understanding this divergence requires addressing a blind spot in securitisation theory. Existing scholarship explains how threats are discursively framed but often neglects what happens after securitisation. The analysis in Section 4 traces this subsequent process. Once AI threats enter specific institutional fields, they are filtered and translated by pre-existing institutional logics. This means a single securitisation input is processed into three distinct governable units. The European Union's legal-bureaucratic logic interprets the AI threat as a potential infringement on legal order and physical well-being. Consequently, it transforms the threat into a *high-risk product* subject to ex ante assessment. The United States' market-liberal logic frames the threat as a manageable risk across sectors. This converts it into a problem of *technical optimisation* requiring continuous stress-testing against performance metrics. China's holistic state logic treats the threat as a challenge to systemic stability. It transforms the threat into a *service responsibility* requiring full-process control through the super-node of *anquan*.

Institutional logics and the dispositif jointly consummate this transformation. The EU's procedure-centric topology results from a legal-bureaucratic logic configured through a dispositif that solidifies AI as a static product locked into a classification-assessment-certification chain. The US sector-dispersed topology reflects a market-liberal logic that pairs AI behaviour with specific harms, governed through decentralized risk assessment and industry self-regulation. China's full-stack delegation topology embodies a holistic state logic, rendering AI as a unit of content generation controlled through service provider accountability and technical value-chain intervention.

While each dispositif is effective within its own logic, their coexistence reveals the essence of structural incommensurability. This refers to ontological differences in governance objects masked by shared vocabulary. In governance studies, cross-system coordination presupposes commensurability, defined as a shared standard of comparison that allows outputs to be translated and mutually recognised (Espeland and Stevens, 1998). Yet, no such conversion rule exists here because the documents measure fundamentally different dimensions, including rights impact, sectoral harm, and order risk. When the EU refers to *AI safety*, it asks whether a technical artefact complies with rights-protection and health standards. When the US discusses *AI risk*, it addresses the probability of system failure and market inefficiency in specific contexts. When China speaks of *security*, it concerns the impact of algorithmic services on social stability. These are ontologically distinct objects rather than merely different preferences regarding the same object. Structural incommensurability is therefore defined here as the ontological divergence of governance objects embedded in institutional logics. This condition renders cross-border coordination operationally impossible to align through mere technical adjustment.

This concept must be distinguished from two adjacent theoretical constructs. First, it differs from Kuhn's (1962) paradigmatic incommensurability, which describes an epistemological state where scientists in different paradigms literally see different worlds. We make no such epistemological claim. Policymakers can read each other's texts and reach rhetorical consensus. The problem is that despite such understanding, incompatibility at the level of the *dispositif* causes alignment failure in execution. Second, it differs from the value incommensurability discussed by Espeland and Stevens (1998). In their analysis, incommensurability often arises from actors' strategic claims that certain values cannot be compared. Our concept describes a different state. Neither China, the EU, nor the US claims that safety is an incomparable value. However, when this term triggers entirely different administrative procedures in each jurisdiction, coordination faces structural barriers. The modifier *structural* carries critical analytical weight. It emphasises that incompatibility stems from the objective characteristics of institutional configurations rather than subjective intent or communication failures. As the data show, even with a sincere desire to coordinate, the divergence in topological structures constitutes a formidable obstacle. Structural incommensurability is thus a state of de facto misalignment where the cost of alignment is structurally prohibitive.

This diagnosis allows for a reassessment of the three existing explanations critiqued in the introduction. Realist accounts attribute fragmentation to power competition, but our findings suggest that jurisdictions are not governing the same object. This renders the premise of competition over a single asset invalid. Institutionalist accounts blame the lack of coordination mechanisms, yet the prerequisite for coordination, namely a shared governable unit, is absent. Instrumentalists point to differences in tool selection, but these differences arise because the tools are applied to fundamentally different objects. Structural incommensurability offers an alternative diagnosis. The barrier lies not in the willingness of parties to coordinate, but in whether existing institutional configurations ontologically permit such coordination.

## 5. Conclusion and Limitation

This study addresses the persistent paradox of rhetorical convergence and operational fragmentation in global AI governance. By integrating securitisation theory with institutional logics through the lens of the dispositif, the analysis demonstrates that this fragmentation arises not from geopolitical rivalry or institutional complexity alone, but from the ontological divergence of the governance objects produced through different pathways of securitisation.

Empirical findings from the comparative semantic network analysis reveal three distinct governance realities. In the European Union, AI is juridified as a static product, where safety is proceduralised through conformity assessments to ensure legal admissibility and physical health protection. In the United States, AI is operationalised as a dynamic engineering system, where safety implies technical robustness and context-specific optimisation against market risks. In China, safety and security are fused within a holistic state logic, rendering AI as socio-technical infrastructure that must be continuously stabilised to align with political order. What's more, these regimes do not merely represent different regulatory styles; they constitute fundamentally different objects of governance. This condition, described here as structural incommensurability, implies that coordination failures stem from the absence of a shared ontological reference point. Consequently, global initiatives assuming equivalence between 'AI safety' regimes risk mistaking semantic overlap for substantive alignment. These finding challenges ethics-by-principles approaches, suggesting that future governance depends less on universal standards than on developing mechanisms of translation capable of operating across divergent regimes.

Several limitations necessitate a cautious interpretation of these findings. Methodologically, while Semantic Network Analysis effectively maps the *topology* of governance discourse, it remains a structuralist approach that may overlook the fluidity of regulatory practice. Policy texts represent the codified ideal of governance, not necessarily its enforcement; thus, a degree of decoupling between the dispositif on paper and its bureaucratic enactment is possible. Future research could usefully complement these topological metrics with qualitative case studies or ethnographic fieldwork to verify whether these semantic divergences persist in daily administrative routines. Empirically, the study is deliberately limited to the 'AI superpowers', potentially overlooking alternative governance models emerging in the Global South or among middle powers. Furthermore, the dataset, restricted to the 2023-2025 regulatory window, captures a specific snapshot of crystallisation. As AI technology evolves, these semantic structures may shift, suggesting that longitudinal analysis over a longer timeframe would be valuable to track the stability of these institutional logics.

Despite these limitations, the study contributes by reframing fragmentation in AI governance as an ontological problem rather than a coordination failure. Governing AI is not simply a matter of aligning instruments, but of recognising how different political systems construct fundamentally different objects under the same name. In this sense, the challenge of global AI governance is not only to agree on what safety means, but to confront the possibility that we may not be governing the same thing at all.

**Reference List**

1) Abbott, K.W., Genschel, P., Snidal, D. and Zangl, B. (eds.) (2017) *The Oxford Handbook of International Law*. Oxford: Oxford University Press.

2) Allen, G.C. and Chan, C. (2023) *Technology Will Remain the Heart of U.S.-China Competition in 2024*. Washington, D.C.: Center for Strategic and International Studies (CSIS).

3) Black, J. (2008) 'Forms and Paradoxes of Principles-Based Regulation', *Capital Markets Law Journal*, 3(4), pp. 425-457.

4) Bradford, A., Aboy, M. and Liddell, K. (2023) 'The Brussels Effect and Artificial Intelligence: How EU Regulation Will Impact the Global AI Market', *European Law Journal*, 29(1-3), pp. 5-28.

5) Buzan, B., Wæver, O. and de Wilde, J. (1998) *Security: A New Framework for Analysis*. Boulder, CO: Lynne Rienner Publishers.

6) Cath, C. (2018) 'Governing Artificial Intelligence: Ethical, Legal and Technical Opportunities and Challenges', *Philosophical Transactions of the Royal Society A*, 376(2133), p. 20180080.

7) Cihon, P., Maas, M.M. and Kemp, L. (2020) 'Fragmentation and the Future: Investigating Architectures for International AI Governance', *Global Policy*, 11(5), pp. 545-556.

8) Cihon, P., Schuett, J. and Baum, S.D. (2021) 'Corporate Governance of Artificial Intelligence in the Public Interest', *Information*, 12(7), p. 275.

9) Creemers, R. (2018) *China's Social Credit System: An Evolving Practice of Control*. Available at: https://ssrn.com/abstract=3175792.

10) Ding, J. (2018) *Deciphering China's AI Dream: The Context, Components, Capabilities, and Consequences of China's Strategy to Lead the World in AI*. Oxford: Future of Humanity Institute, University of Oxford.

11) Espeland, W.N. and Stevens, M.L. (1998) 'Commensuration as a Social Process', *Annual Review of Sociology*, 24, pp. 313-343.

12) Floridi, L., Cowls, J., Beltrametti, M., Chatila, R., Chazerand, P., Dignum, V., *et al.* (2018) 'AI4People—An Ethical Framework for a Good AI Society: Opportunities, Risks, Principles, and Recommendations', *Minds and Machines*, 28(4), pp. 689-707.

13) Foucault, M. (1980) 'The Confessions of the Flesh', in Gordon, C. (ed.) *Power/Knowledge: Selected Interviews and Other Writings 1972-1977*. New York: Pantheon Books, pp. 194-228.

14) Friedland, R. and Alford, R.R. (1991) 'Bringing Society Back In: Symbols, Practices, and Institutional Contradictions', in Powell, W.W. and DiMaggio, P.J. (eds.) *The New Institutionalism in Organizational Analysis*. Chicago: University of Chicago Press, pp. 232-263.

15) Gordon, N., Rieder, G. and Sileno, G. (2022) 'On Mapping Values in AI Governance', *Computer Law & Security Review*, 46, p. 105736.

16) Gunningham, N. (2021) 'A New Regulatory Paradigm for Artificial Intelligence: Self-Regulation, Co-Regulation or Meta-Regulation?', *Regulation & Governance*, 17(1), pp. 307-323.

17) Horowitz, M.C. (2018) 'Artificial Intelligence, International Competition, and the Balance of Power', *Texas National Security Review*, 1(3), pp. 37-57.

18) Kroll, J.A., Huey, J., Barocas, S., Felten, E.W., Reidenberg, J.R., Robinson, D.G. and Yu, H. (2017) 'Accountable Algorithms', *University of Pennsylvania Law Review*, 165(3), pp. 633-705.

19) Kuhn, T.S. (1962) *The Structure of Scientific Revolutions*. Chicago: University of Chicago Press.

20) Luna, J., Tan, I., Xie, X. and Jiang, L. (2024) 'Navigating Governance Paradigms: A Cross-Regional Comparative Study of Generative AI

Governance', in *Proceedings of the 2024 AAAI/ACM Conference on AI, Ethics, and Society (AIES '24)*. New York: Association for Computing Machinery.

21) Marsden, C. and Meyer, T. (2023) *Regulating Disinformation with Artificial Intelligence: Effects of Disinformation Initiatives on Freedom of Expression and Media Freedom*. Strasbourg: Council of Europe.

22) National Institute of Standards and Technology (NIST) (2023) *AI Risk Management Framework (AI RMF 1.0)*. Gaithersburg, MD: U.S. Department of Commerce.

23) Raustiala, K. and Victor, D.G. (2004) 'The Regime Complex for Plant Genetic Resources', *International Organization*, 58(2), pp. 277-309.

24) Roberts, H., Triolo, P. and Ferguson, E. (2021) *Understanding Chinese Government Guidance Funds: An Analysis of Chinese-Language Policy Documents*. Washington, D.C.: Center for Security and Emerging Technology (CSET).

25) Rose, N. and Miller, P. (1990) 'Governing Economic Life', *Economy and Society*, 19(1), pp. 1-31.

26) Rose, N. and Miller, P. (1992) 'Political Power Beyond the State: Problematics of Government', *The British Journal of Sociology*, 43(2), pp. 173-205.

27) Rose, N. and Miller, P. (2008) *Governing the Present: Administering Economic, Social and Personal Life*. Cambridge: Polity Press.

28) Schmitt, L. (2022) 'Mapping Global AI Governance: A Nascent Regime Complex', *AI and Ethics*, 2(2), pp. 303-314.

29) Thornton, P.H., Ocasio, W. and Lounsbury, M. (2012) *The Institutional Logics Perspective: A New Approach to Culture, Structure, and Process*. Oxford: Oxford University Press.

30) Veale, M. and Borgesius, F.Z. (2021) 'Demystifying the Draft EU Artificial Intelligence Act—Analysing the Good, the Bad, and the Unclear Elements of the Proposed Approach', *Computer Law Review International*, 22(4), pp. 97-112.

## Appendix A: Detailed Topological Metrics

The following tables present the complete centrality metrics for the top nodes in each jurisdiction's semantic network. The metrics include Frequency, Degree Centrality, Betweenness Centrality, and Eigenvector Centrality.

Table A1: Comprehensive Topological Metrics of the EU AI Governance Corpus (2023–2025)

| Term | Frequency | Degree_Centrality | Betweenness | Eigenvector |
|---|---|---|---|---|
| high_risk | 440 | 0.3673 | 0.0952 | 0.2254 |
| ai_system | 1207 | 0.3265 | 0.0102 | 0.045 |
| right | 195 | 0.3061 | 0.023 | 0.0147 |
| fundamental_rights | 135 | 0.3061 | 0.1037 | 0.0298 |
| trustworthy_ai | 106 | 0.2857 | 0.1497 | 0.0381 |
| safety | 154 | 0.2857 | 0.0927 | 0.0522 |
| risk | 320 | 0.2653 | 0 | 0.0253 |
| general_purpose_ai | 216 | 0.2653 | 0.0561 | 0.053 |
| human | 176 | 0.2653 | 0.0918 | 0.0215 |
| security | 93 | 0.2449 | 0.0425 | 0.0395 |
| systemic_risk | 132 | 0.2449 | 0.1378 | 0.0672 |
| service | 195 | 0.2245 | 0.0068 | 0.0678 |
| person | 314 | 0.2245 | 0.0043 | 0.009 |
| conformity_assessment | 77 | 0.1837 | 0.2602 | 0.4546 |
| compliance | 147 | 0.1837 | 0 | 0.0418 |
| robustness | 31 | 0.1429 | 0.216 | 0.0457 |
| documentation | 84 | 0.1429 | 0.0179 | 0.2426 |
| principle | 102 | 0.1429 | 0.0153 | 0.021 |
| value | 48 | 0.1224 | 0.0179 | 0.022 |
| notified_body | 66 | 0.1224 | 0.0408 | 0.4513 |
| freedom | 42 | 0.1224 | 0.0884 | 0.0115 |
| performance | 34 | 0.1224 | 0.2211 | 0.0761 |
| product | 108 | 0.1224 | 0.0153 | 0.1473 |
| guideline | 44 | 0.102 | 0.0102 | 0.0239 |
| input | 34 | 0.102 | 0.0714 | 0.0357 |
| health | 83 | 0.102 | 0.1173 | 0.0444 |
| quality_management | 31 | 0.102 | 0.023 | 0.4505 |
| biometric_identification | 43 | 0.0816 | 0.0136 | 0.0053 |
| report | 136 | 0.0816 | 0.0119 | 0.0158 |
| personal_data | 66 | 0.0816 | 0.0289 | 0.0038 |
| deployment | 36 | 0.0816 | 0 | 0.0144 |
| technical_documentation | 45 | 0.0816 | 0.0068 | 0.3986 |
| transparency | 46 | 0.0816 | 0.0017 | 0.0153 |
| environment | 32 | 0.0816 | 0.0332 | 0.0164 |
| user | 46 | 0.0816 | 0.0306 | 0.0123 |
| framework | 107 | 0.0816 | 0 | 0.0156 |
| legislation | 52 | 0.0816 | 0.0323 | 0.1908 |
| individual | 89 | 0.0816 | 0.0119 | 0.0076 |
| harm | 68 | 0.0816 | 0.0238 | 0.0078 |
| accuracy | 29 | 0.0612 | 0.0111 | 0.0344 |
| placing_on_the_market | 34 | 0.0612 | 0.023 | 0.0878 |
| output | 52 | 0.0612 | 0.0459 | 0.0181 |
| standard | 83 | 0.0612 | 0.0077 | 0.0173 |
| oversight | 47 | 0.0612 | 0.0485 | 0.0164 |
| criterion | 33 | 0.0612 | 0.0544 | 0.0811 |
| market_surveillance | 135 | 0.0408 | 0.0051 | 0.0335 |
| bias | 48 | 0.0408 | 0.0272 | 0.0024 |
| consumer | 29 | 0.0408 | 0.0017 | 0.0023 |
| responsibility | 52 | 0.0204 | 0 | 0.0642 |

Table A2: Comprehensive Topological Metrics of the United States AI Governance Corpus (2023–2025)

| Term | Frequency | Degree_Centrality | Betweenness | Eigenvector |
|---|---|---|---|---|
| risk | 152 | 0.3878 | 0.0587 | 0.0145 |
| right | 105 | 0.3673 | 0.0451 | 0.034 |
| ai | 151 | 0.3673 | 0.0408 | 0.0105 |
| privacy | 72 | 0.3469 | 0.0255 | 0.0517 |
| automated_system | 209 | 0.3265 | 0.0247 | 0.0156 |
| harm | 72 | 0.3061 | 0.0051 | 0.0016 |
| bias | 30 | 0.2857 | 0.0459 | 0.0171 |
| safety | 26 | 0.2857 | 0.0502 | 0.0049 |
| ai_system | 69 | 0.2653 | 0.0451 | 0.015 |
| discrimination | 49 | 0.2449 | 0.0595 | 0.0147 |
| individual | 63 | 0.2449 | 0.0255 | 0.0109 |
| security | 25 | 0.2449 | 0.0162 | 0.0134 |
| people | 63 | 0.2245 | 0.0043 | 0.0211 |
| employment | 16 | 0.1837 | 0.0595 | 0.4014 |
| principle | 82 | 0.1837 | 0.051 | 0.0088 |
| surveillance | 42 | 0.1837 | 0.0825 | 0.2828 |
| finance | 6 | 0.1633 | 0.0791 | 0.401 |
| high_risk | 8 | 0.1633 | 0.1131 | 0.1277 |
| health | 21 | 0.1633 | 0.0068 | 0.3566 |
| framework | 51 | 0.1633 | 0.0009 | 0.0046 |
| risk_assessment | 6 | 0.1633 | 0.1769 | 0.0684 |
| community | 53 | 0.1633 | 0.0026 | 0.0014 |
| education | 16 | 0.1633 | 0.0519 | 0.437 |
| standard | 42 | 0.1633 | 0.0077 | 0.0022 |
| oversight | 27 | 0.1633 | 0.0672 | 0.1362 |
| risk_management | 26 | 0.1429 | 0.0502 | 0.0159 |
| environment | 8 | 0.1429 | 0.0238 | 0.0117 |
| personal_data | 6 | 0.1429 | 0.0714 | 0.1515 |
| responsibility | 11 | 0.1429 | 0.068 | 0.015 |
| trustworthy | 7 | 0.1429 | 0.0204 | 0.0102 |
| justice | 18 | 0.1429 | 0.0731 | 0.3851 |
| service | 23 | 0.1224 | 0.0357 | 0.1277 |
| consumer | 14 | 0.1224 | 0 | 0.175 |
| report | 55 | 0.1224 | 0.0255 | 0.0051 |
| algorithm | 23 | 0.1224 | 0.0595 | 0.0234 |
| record | 13 | 0.102 | 0.0332 | 0.0158 |
| product | 12 | 0.102 | 0.1207 | 0.0366 |
| human | 82 | 0.102 | 0 | 0.0164 |
| transparency | 10 | 0.102 | 0.0646 | 0.0373 |
| output | 8 | 0.0816 | 0.074 | 0.0135 |
| performance | 16 | 0.0816 | 0.0383 | 0.003 |
| person | 20 | 0.0816 | 0.0068 | 0.0983 |
| deployment | 35 | 0.0816 | 0.0026 | 0.0015 |
| machine_learning | 8 | 0.0816 | 0.0434 | 0.0055 |
| decision_making | 8 | 0.0816 | 0.017 | 0.0176 |
| guideline | 7 | 0.0612 | 0.0689 | 0.0389 |
| documentation | 12 | 0.0408 | 0.0179 | 0.0045 |
| value | 10 | 0.0408 | 0 | 0.004 |
| user | 46 | 0.0408 | 0 | 0.0007 |

Table A3: Comprehensive Topological Metrics of the Chinese AI Governance Corpus (2023–2025)

| Term | Term in English | Frequency | Degree_Centrality | Betweenness | Eigenvector |
|---|---|---|---|---|---|
| 安全 ānquán | safety | 196 | 0.2857 | 0.051 | 0.0142 |
| 人工智能 réngōng zhìnéng | artificial intelligence | 371 | 0.2449 | 0 | 0.0031 |
| 模型 móxíng | model | 102 | 0.2245 | 0.1607 | 0.026 |
| 研发 yánfā | R&D | 27 | 0.1633 | 0.2976 | 0.0848 |
| 探索 tànsuǒ | exploration | 21 | 0.1633 | 0.0655 | 0.009 |
| 服务提供者 fúwù tígōngzhě | service provider | 24 | 0.1633 | 0.1692 | 0.3445 |
| 训练 xùnliàn | training | 36 | 0.1429 | 0.0978 | 0.0224 |
| 措施 cuòshī | measures | 27 | 0.1429 | 0.1335 | 0.1113 |
| 算法 suànfǎ | algorithm | 37 | 0.1429 | 0.2857 | 0.0338 |
| 分级 fēnjí | classification | 25 | 0.1429 | 0.1276 | 0.0443 |
| 生成合成 shēngchéng héchéng | generation/synthesis | 34 | 0.1429 | 0.0578 | 0.4843 |
| 内容 nèiróng | content | 73 | 0.1224 | 0.0425 | 0.4211 |
| 加快 jiākuài | accelerate | 22 | 0.1224 | 0.0043 | 0.0055 |
| 协同 xiétóng | collaboration | 21 | 0.1224 | 0.0893 | 0.0106 |
| 创新 chuàngxīn | innovation | 25 | 0.1224 | 0.1003 | 0.011 |
| 添加 tiānjiā | add | 24 | 0.1224 | 0.0281 | 0.4022 |
| 数据 shùjù | data | 89 | 0.1224 | 0.04 | 0.0476 |
| 采取 cǎiqǔ | adopt | 18 | 0.1224 | 0.1029 | 0.2259 |
| 系统 xìtǒng | system | 48 | 0.1224 | 0.0187 | 0.0018 |
| 技术 jìshù | technology | 83 | 0.1224 | 0.0043 | 0.0327 |
| 治理 zhìlǐ | governance) | 44 | 0.1224 | 0.0663 | 0.0027 |
| 输出 shūchū | output) | 30 | 0.102 | 0.0204 | 0.0083 |
| 决策 juécè | decision-making | 24 | 0.102 | 0.0791 | 0.0053 |
| 场景 chǎngjǐng | scenario | 42 | 0.102 | 0.0128 | 0.0568 |
| 应对 yìngduì | response | 32 | 0.102 | 0.0196 | 0.0434 |
| 伦理 lúnlǐ | ethics | 19 | 0.102 | 0.057 | 0.0048 |
| 社会 shèhuì | society | 43 | 0.102 | 0.0383 | 0.0019 |
| 标识 biāozhì | identifier | 58 | 0.102 | 0 | 0.4597 |
| 服务 fúwù | service | 68 | 0.0816 | 0.0408 | 0.0791 |
| 过程 guòchéng | process | 19 | 0.0816 | 0 | 0.0126 |
| 攻击 gōngjī | attack | 20 | 0.0816 | 0.0068 | 0.0045 |
| 发展 fāzhǎn | development | 59 | 0.0816 | 0.0204 | 0.0031 |
| 形成 xíngchéng | formation | 17 | 0.0816 | 0 | 0.0028 |
| 信息 xìnxī | information | 59 | 0.0816 | 0 | 0.0669 |
| 建设 jiànshè | construction | 29 | 0.0816 | 0.0017 | 0.0027 |
| 防范 fángfàn | prevention | 25 | 0.0816 | 0.0408 | 0.0253 |
| 智能化 zhìnéng huà | intelligentization | 21 | 0.0816 | 0.0077 | 0.0141 |
| 确保 quèbǎo | ensure | 20 | 0.0612 | 0.0009 | 0.0026 |
| 国家 guójiā | country | 18 | 0.0612 | 0.0315 | 0.0005 |
| 开源 kāiyuán | open source | 27 | 0.0612 | 0.0391 | 0.0038 |
| 影响 yǐngxiǎng | impact | 33 | 0.0612 | 0.0408 | 0.0009 |
| 制定 zhìdìng | formulate | 18 | 0.0612 | 0 | 0.0113 |
| 强化 qiánghuà | strengthen | 21 | 0.0408 | 0 | 0.0006 |
| 分类 fēnlèi | classification | 20 | 0.0408 | 0 | 0.0157 |
| 机制 jīzhì | mechanism | 27 | 0.0408 | 0 | 0.0011 |
| 全球 quánqiú | global | 19 | 0.0408 | 0.0281 | 0.0007 |
| 资源 zīyuán | resource | 18 | 0.0408 | 0.0085 | 0.0012 |
| 规范 guīfàn | standard | 20 | 0.0408 | 0.0068 | 0.0033 |
| 产品 chǎnpǐn | product | 22 | 0.0204 | 0 | 0.0128 |

## Appendix B: Mathematical Formulations of Topological Metrics

To ensure reproducibility and methodological transparency, this appendix details the mathematical formulations of the network metrics employed in the analysis. All metrics were computed using Python's NetworkX library.

### B. Normalised Pointwise Mutual Information (NPMI)

Edges in the semantic network are constructed and weighted based on NPMI to filter out trivial co-occurrences. NPMI normalises the Pointwise Mutual Information (PMI) to a range of $[-1, 1]$ where positive values indicate a semantic association greater than chance.

The standard PMI between two terms, $x$ and $y$, is defined as

$$PMI(x,y) = \log_2 \frac{P(x,y)}{P(x)P(y)} \quad (1)$$

Where:

- $P(x)$ and $P(y)$ are the probabilities of terms $x$ and $y$ occurring in the corpus (number of windows containing the term divided by the total number of windows).
- $P(x,y)$ is the probability of terms $x$ and $y$ co-occurring in the same sliding window.

To standardise the measure for comparison across pairs with varying frequencies, NPMI is calculated as:

$$NPMI(x,y) = \frac{PMI(x,y)}{h(x,y)} \quad (2)$$

Where $h(x,y)$ is the self-information (or negative log probability) of the cooccurrence:

$$h(x,y) = -\log_2 P(x,y) \quad (3)$$

In this study, edges with $NPMI < 0$ were pruned to remove negative associations, ensuring the network topology reflects only positive semantic bindings.

## Generative AI Usage Statement

Generative AI tools were **not** used for content generation or writing in this paper. All substantive content, including theoretical arguments, empirical analysis, and interpretations, was authored entirely by us, the human authors. Generative AIs were only used solely for: (1) formatting review and proofreading assistance, and (2) soliciting preliminary feedback on structural organisation during the revision process. No AI-generated text appears in the final manuscript. The authors bear full responsibility for the originality, accuracy, and integrity of all content presented herein.